\documentclass[aps,prl,reprint]{revtex4-1}

\usepackage{graphicx}
\usepackage{subfigure}
\usepackage{color}
\usepackage{amsmath}
\usepackage[linktocpage,colorlinks=true,linkcolor=blue,citecolor=blue,breaklinks=true,urlcolor=blue]{hyperref}
\usepackage{verbatim}
\usepackage{breakcites}
\usepackage{wrapfig}
\usepackage{soul}
\usepackage{ulem}

\begin{document}

\title{Upstream swimming in microbiological flows}

\author{Arnold J. T. M. Mathijssen}
\affiliation{The Rudolf Peierls Centre for Theoretical Physics, 1 Keble Road, Oxford, OX1 3NP, UK}
\email[Correspondence: ]{mathijssen@physics.ox.ac.uk}

\author{Tyler N. Shendruk}
\affiliation{The Rudolf Peierls Centre for Theoretical Physics, 1 Keble Road, Oxford, OX1 3NP, UK}

\author{Julia M. Yeomans}
\affiliation{The Rudolf Peierls Centre for Theoretical Physics, 1 Keble Road, Oxford, OX1 3NP, UK}

\author{Amin Doostmohammadi}
\affiliation{The Rudolf Peierls Centre for Theoretical Physics, 1 Keble Road, Oxford, OX1 3NP, UK}

\newcommand{\cut}[1]{\sout{#1}}
\newcommand{\add}[1]{\textit{#1}}

\date{\today}

\begin{abstract}
\noindent
Interactions between microorganisms and their complex flowing environments are essential in many biological systems. We develop a model for microswimmer dynamics in non-Newtonian Poiseuille flows. We predict that swimmers in shear-thickening (-thinning) fluids migrate upstream more (less) quickly than in Newtonian fluids and demonstrate that viscoelastic normal stress differences reorient swimmers causing them to migrate upstream at the centreline, in contrast to well-known boundary accumulation in quiescent Newtonian fluids. Based on these observations, we suggest a sorting mechanism to select microbes by swimming speed.
\end{abstract}


\maketitle

\begin{par}
Motile microorganisms ubiquitously inhabit confined and complex microenvironments.
Geometrical constraints are a key regulator of rheotaxis, the reorientation of swimmers in response to external flows \cite{miki2013}, and are essential in the design of microfluidic devices for drug delivery systems, hematology and cytometry \cite{Simone2008,Brown2000}. 
Additionally, the complexity of embedding fluids is crucial. One important aspect of the complexity arises from the dual fluidic and elastic (viscoelastic) behaviour of many biological fluids such as mucus and extracellular matrix gels~\cite{Berger2000,Hwang1969,Gilboa1976} or blood at macroscopic length-scales \cite{moriarty2008real, hickey2009intravascular, radolf2012ticks}.
\end{par}
\begin{par}
Important correlations have been found between non-Newtonian behaviour of the fluid and pathological phenomena. 
Gastric mucus viscoelasticity effects swimming of {\it H. pylori}, an abundant pathogen in the stomach and leading cause of ulcers \cite{montecucco2001living, Celli2009}. It has been shown that viscoelasticity is a more crucial factor in controlling the maximum velocity of lyme disease pathogen {\it B. burgdorferi} through skin than even chemical composition \cite{Kimsey1990}. 
Viscoelastic properties of mucus have a remarkable impact on the swimming of spermatozoa and sperm-egg encounter rates \cite{Riffell2007}. 
 \end{par}
\begin{par}
Despite the widespread implications of viscoelastic effects on biological processes, research on motile microorganism dynamics in confined environments is largely limited to Newtonian fluids \cite{karimi2013hydrodynamic, uppaluri2012flow, zottl2012nonlinear, costanzo2012transport, zottl2013periodic, chacon2013chaotic, ledesma2013enhanced, rusconi2014bacterial, chilukuri2014impact, costanzo2014motility, tung2015emergence}. Recently, a large number of studies have considered locomotion in quiescent non-Newtonian fluids at the scale of microswimmers, in experiment, simulations and theory\cite{liu2011force, gagnon2014undulatory, gao2014altered, martinez2014flagellated,lauga2007propulsion, fu2010low, teran2010viscoelastic, spagnolie2013locomotion},
but little is known about the dynamical behaviour of swimmers subject to large-scale non-Newtonian flows. 
\end{par}
\begin{par}
In this work, we construct a tractable theoretical framework for individual microorganisms swimming in confined, flowing microbiological environments of non-Newtonian fluids. We study the macroscopic effects of shear-dependent viscosity and viscoelasticity, both in separation and in conjunction, for a weakly viscoelastic fluid. Image systems are introduced, regularizing the hydrodynamic interaction of microswimmers with the walls, and swimmer trajectories are characterized. Shear-dependent viscosity is seen to greatly impact the upstream motion of motile cells and our analysis shows that the presence of normal stress differences in viscoelastic fluids results in a remarkable upstream migration along the centreline. 
We provide quantitative measures of the upstream motion and propose a novel sorting mechanism for motile organisms in confined viscoelastic flows.
\end{par}

\begin{figure}[b!]
	\begin{center}
	\includegraphics[width=0.8 \linewidth]{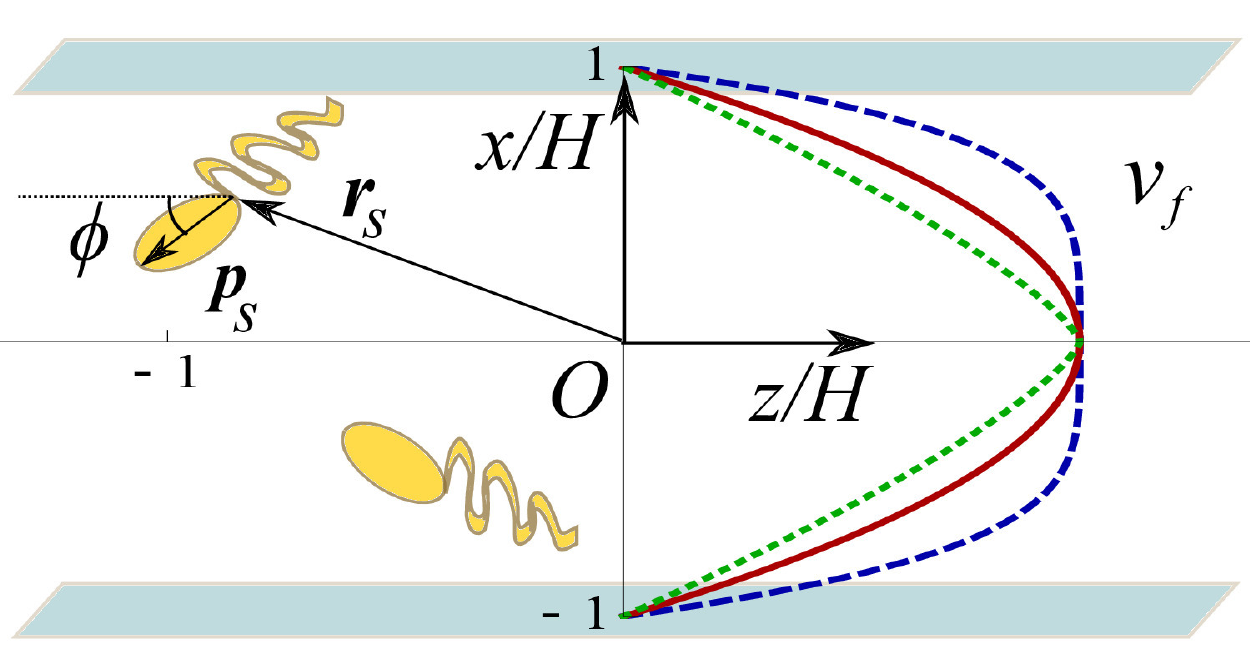}
	\caption{Schematic of a microswimmer at position $\boldsymbol{r}_\textmd{s}$ and moving with speed $v_\textmd{s}$ in the direction $\boldsymbol{p}_\textmd{s}$ subject to a viscoelastic flow within a microchannel of height $2H$. The Poiseuille flow $\boldsymbol{v}_\textmd{f}$ is shown for shear-thinning (blue, dashed), Newtonian (red, solid) and shear-thickening (green, dotted) fluids.}
	\label{fig:GeometryDiagram}
	\end{center}
\end{figure}
\begin{par}
A single microorganism is modeled as swimming in flowing, incompressible, non-Newtonian fluid within a channel of height $2H$ (Fig.~\ref{fig:GeometryDiagram}). In addition to its swimming velocity $\boldsymbol{v}_\textmd{s}=v_\textmd{s} \boldsymbol{p}_\textmd{s}$ in the direction $\boldsymbol{p}_\textmd{s}$, the motion of the swimming cell of radius $a$ is affected by the background flow $\boldsymbol{v}_\textmd{f}$, hydrodynamic interactions (HI) with the channel walls $\boldsymbol{v}_\textmd{HI}$, and cross-streamline migration induced by viscoelastic normal stress difference gradients $\boldsymbol{v}_\textmd{M}$. Thus, the evolution of a microswimmer's position and direction are
\begin{eqnarray}
 \label{eq:EquationsOfMotion}
 \dot{\boldsymbol{r}}_\textmd{s} &=& \boldsymbol{v}_\textmd{s} + \boldsymbol{v}_\textmd{f} + \boldsymbol{v}_\textmd{HI} + \boldsymbol{v}_\textmd{M} \\
\label{eq:EquationsOfMotion2}
\dot{\boldsymbol{p}}_\textmd{s} &=& \boldsymbol{\Omega}_\textmd{f} \times\boldsymbol{p}_\textmd{s}
 + \boldsymbol{\Omega}_\textmd{HI} \times \boldsymbol{p}_\textmd{s},
\end{eqnarray}
where $\boldsymbol{\Omega}_\textmd{f} =  \frac{1}{2} \boldsymbol{\nabla} \times \boldsymbol{v}_\textmd{f}$ and $\boldsymbol{\Omega}_\textmd{HI}$ denotes the angular velocity due to the HI with the walls.

The translational invariance of Eqs.~(\ref{eq:EquationsOfMotion}-\ref{eq:EquationsOfMotion2}) along the $y$ and $z$ directions allows us to consider motion of swimmers in the $y=0$ plane and orientation can be represented in cylindrical coordinates as $\boldsymbol{p}_\textmd{s} = - \sin(\phi)\boldsymbol{\hat{e}}_x - \cos(\phi)\boldsymbol{\hat{e}}_z$, where $\phi \in [-\pi, \pi]$ is the angle in the $x-z$ plane. Upstream swimming corresponds to $\phi=0$ and downstream to $\pm\pi$ (Fig.~\ref{fig:GeometryDiagram}). Consequently, the dynamics of the system can be represented by two coupled equations, $\dot{x} = \dot{x}(x,\phi)$ and $\dot{\phi} = \dot{\phi}(x,\phi)$, and a third uncoupled equation $\dot{z} = \dot{z}(x,\phi)$. 
We nondimensionalise lengths by half the channel height, $H$, and velocities by the swimming speed, $v_\textmd{s}$. Therefore, changes in the swimming speed due to viscoelasticity, as studied in Refs.
\cite{liu2011force, gagnon2014undulatory, gao2014altered, martinez2014flagellated, lauga2007propulsion, fu2010low, teran2010viscoelastic, spagnolie2013locomotion}, 
are readily incorporated in this model.
\end{par}

\begin{figure}[b]
	\begin{center}
	\includegraphics[trim=25mm 220mm 115mm 18.8mm, clip=true, width=\linewidth]{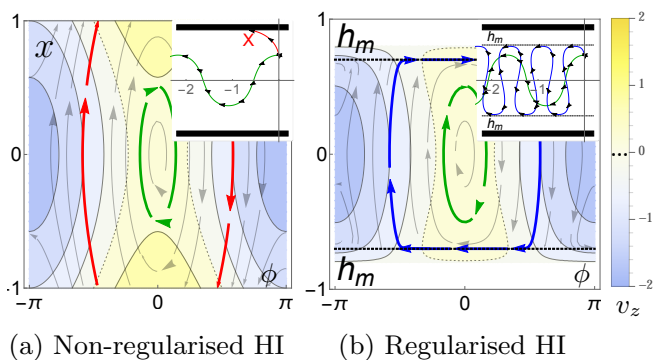}
	\caption{Typical trajectories for swimmer dynamics in a Newtonian Poiseuille flow shown in $x-\phi$ phase space, and in the $x-z$ plane (insets).
The swimmer and maximum flow velocities are $v_\textmd{s} = 1$ and $v_\textmd{max} = 0.75$, and the dipole moment is $\kappa = 0$. 
a) Non-regularised HI, $\sigma = 0$. 
b) Regularised HI with $\sigma = (3/10)^{3}$ so that $h_m = 3/10$. 
The background colours indicate the velocity in the $z$ direction.}
        \label{fig:ResultsNewtonian}
	\end{center}
\end{figure}

\begin{par}
In a Newtonian fluid, this system shows the emergence of swinging and tumbling microswimmer trajectories in Poiseuille flow~\cite{zottl2012nonlinear, zottl2013periodic}. 
Upstream-oriented swimmers are rotated by background vorticity so that they oscillate about the centreline (Fig. \ref{fig:ResultsNewtonian}(a-b); green trajectory). 
For large oscillation amplitudes, however, the swimmer runs into the walls (Fig.~\ref{fig:ResultsNewtonian}(a); red trajectory).
Hence, HI with the boundaries must be included \cite{zottl2012nonlinear}. 
Simply including the far-field force dipole of strength $\kappa$ and an image system consisting of a superposition of point-force singularities \cite{spagnolie2012hydrodynamics} in the HI produces non-physical singular flow fields near the walls, unless a physical cut-off length is provided.

We construct a more physical representation by including a source doublet of strength $\sigma$ in the swimmer's flow and image fields, producing a more accurate near-field flow and regularising the HI with the boundaries. 
This ensures that the swimmer is turned away from the boundaries by the closest distance of approach $h_m = (\sigma/v_\textmd{s})^{1/3}$, which sets a natural cut-off and gives an effective size.
This may be understood to be its hydrodynamic radius, $a_h  = (2 \sigma/v_\textmd{s})^{1/3}$~\cite{mathijssen2015tracer}, which we expect to be directly proportional to the swimmer size; $a_h \sim a$, and thus $h_m \sim a$. 
E.g. \textit{Volvox} has $\sigma \sim 10^9 \mu\textmd{m}^4\textmd{/s}$ and $v_s \sim 10^2 \mu\textmd{m/s}$  \cite{drescher2010direct}, so that $a_h \sim 270 \mu\textmd{m}$ compared to $a \sim 200 \mu$m.
By including the near-field correction, unphysical swimmer-wall contact is ruled out and the swimmer trajectory runs parallel to the wall with the offset $h_m$ (Fig.~\ref{fig:ResultsNewtonian}(b); blue trajectory). 
To consistently account for finite size effects, we also include the Fax\'{e}n corrections to the flow induced translational, $\boldsymbol{v}_\textmd{f}$, and angular velocity, $\boldsymbol{\Omega}_\textmd{f}$, of the swimmer \cite{kimmicrohydrodynamics}.
\end{par}

\begin{par}
Non-Newtonian effects modify the background flow and trajectories of microswimmers. Non-Newtonian fluids generally feature two properties different from a Newtonian counterpart --- namely, shear dependent viscosity and normal stress differences. Here shear-thinning and -thickening effects are accounted for via a power-law fluid model $\eta = \eta_0 (\dot{\gamma}/\dot{\gamma}_0)^{n-1}$,
where $\dot{\gamma}$ is the shear rate, $\eta_0$ is the viscosity at the shear-rate $\dot{\gamma}_0$, and $n$ is the shear-thinning parameter. 
The background Poiseuille flow of a power-law fluid is
\begin{eqnarray}
\label{eq:TextPoiseuilleFlow}
\boldsymbol{v}_\textmd{f}(\boldsymbol{r}) &=& v_\textmd{max} \left(1-\left(|x|/H \right)^\frac{1+n}{n} \right) \boldsymbol{\hat{e}}_z, \quad
\end{eqnarray}
where $v_\textmd{max}$ is the maximum flow speed. This results in a stronger (weaker) flow near the walls, in shear-thinning (-thickening) fluids compared to a Newtonian fluid with the same $v_\textmd{max}$ (Fig.~\ref{fig:GeometryDiagram}). 
HI with the walls remain approximately Newtonian for weakly non-Newtonian fluids since the asymmetric correction for a dipolar swimmer~\cite{pak2012micropropulsion, zhu2012self} decays rapidly as $\sim r^{-3}$ \cite{acharya76,phillips07}, which is small compared to the Newtonian contribution and amounts to a minor correction on the quadrupolar term.
\end{par}

\begin{figure}[b]
	\includegraphics[trim=2cm 176mm 11cm 1.8cm, clip=true, width=\linewidth]{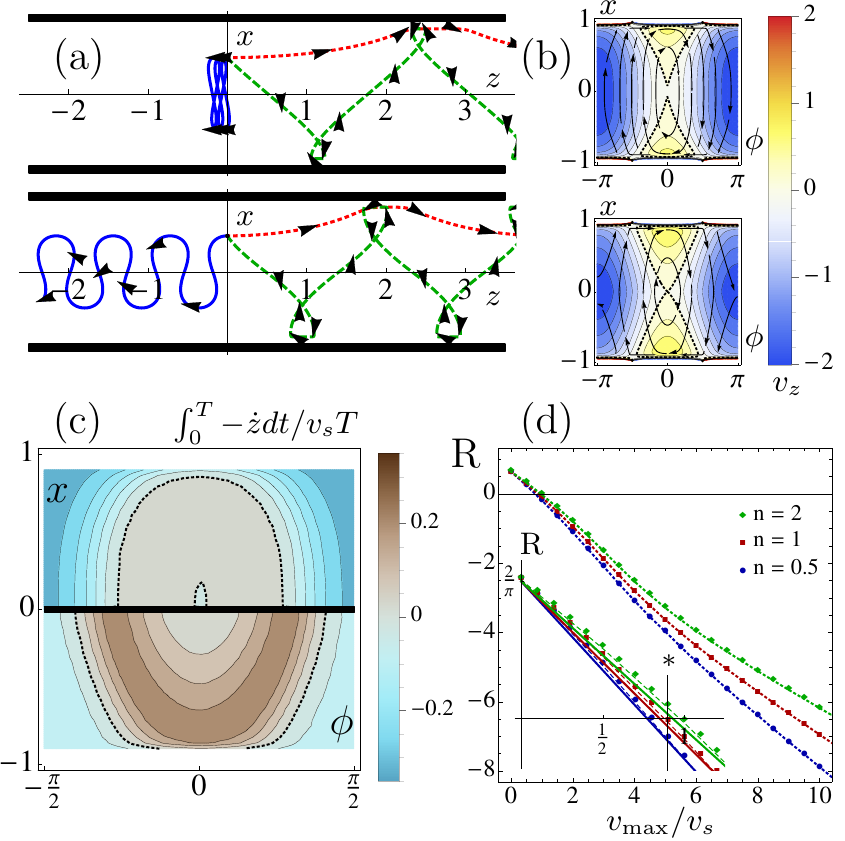}
        \caption{Swimmer dynamics in Poiseuille flow of a shear-thinning ($n=\tfrac{1}{2}$) and -thickening fluid ($n=2$) without normal stresses, shown in the upper and lower halves of subfigures (a-c). 
$v_\textmd{s} = v_\textmd{max} = 1$, $\kappa = 0$ and $\sigma = (1/10)^{3}$.
a) Trajectories in the $x-z$ plane, with initial position $\boldsymbol{r}_s=(0, 1/2)$ and orientations $\phi = 0$ (blue), $\pi/2$ (green), and $\pi$ (red).
b) Trajectories in $x-\phi$ phase space. The background colours indicate the velocity in the $z$ direction.
c) Upstream swimming velocity, $-\dot{z}$, averaged over a large time, for all upstream-oriented initialisations in $x-\phi$ space. 
d) Upstream retention ratio, defined by (c) averaged over these initial conditions, as a function of the flow speed. Points show full numerical solutions, dashed lines show theoretical predictions, and solid lines show the limit $v_\textmd{max} \ll v_\textmd{s}$, Eq.\eqref{eq:RetentionRatio}. The inset focusses on this limit. 
}
        \label{fig:ResultsShear}
\end{figure}

\begin{par}
The upstream motion of a swimmer is enhanced in a shear-thickening fluid compared to a shear-thinning counterpart without normal stresses (Fig.~\ref{fig:ResultsShear} and Supplemental Material movie 1~\cite{Movies}). This is associated with changes in vorticity in the vicinity of the walls. The stronger vorticity of the shear-thinning fluid near the wall results in a more rapid reorientation towards the centreline. 
Consequently, the swimmer has less time to move upstream. 
\end{par}

\begin{par}
An initially upstream oriented swimmer (Fig.~\ref{fig:ResultsShear}(a); blue trajectory) in a shear-thinning fluid moves a short distance upstream after the first oscillation about the centreline, whereas the swimmer in the shear-thickening fluid progresses an order of magnitude further. 
Swimmers initially orientated towards the walls (dashed green trajectories) are carried by the flow, but in a shear-thickening fluid they move further upstream near the walls. 
Similarly, swimmers initially orientated downstream (dotted red trajectories) experience an enhanced downstream motion in a shear-thinning fluid.
This demonstrates that the dynamics in flowing non-Newtonian environments can have a more significant effect on motion than relatively small modifications to the swimming speed in quiescent non-Newtonian fluids \cite{liu2011force, gagnon2014undulatory, gao2014altered, martinez2014flagellated, lauga2007propulsion, fu2010low, teran2010viscoelastic, spagnolie2013locomotion}.
\end{par}

\begin{par}
If $v_\textmd{max} = v_\textmd{s}$, swimmers oriented directly upstream at the centreline do not progress, while those that oscillate about the centreline experience less counterflow on average and therefore are able to migrate upstream (Fig.~\ref{fig:ResultsShear}(b-c)). 
However, if the oscillations about the centreline are too large, the swimmer cannot move upstream. 
Therefore, the effective upstream motility is not well described by any given trajectory but rather by a retention ratio \cite{Wahlund201397}, the ratio of the time-averaged $z$-component of swimmer velocity to the swimming speed, $R = \left\langle T^{-1} \int_0^T - \dot{z}(t; x_0, \phi_0) dt \right\rangle / v_\textmd{s}$,
where we average over all upstream-oriented trajectories $x_0 \in [-H+h_m, H-h_m]$ and $\phi_0 \in [-\pi/2, \pi/2]$.
The upstream retention ratio can be determined numerically (Fig.~\ref{fig:ResultsShear}(d); points) and be approximated analytically. A conserved quantity of motion can be found by integrating $\dot{x}/\dot{\phi}$, giving $\mathcal{C} = 1+ \tfrac{1}{2} v_\text{max} |x|^{(1+n)/n} - v_\textmd{s} \cos \phi$. Hence, the distance travelled along $z$ per oscillation can be computed, $D = \int_{traj} \dot{z} dt$, as well as the period, $T = \int_{traj} dt$. Dividing these and averaging over the initial conditions gives $R = - \langle \tfrac{D}{v_\textmd{s}T} \rangle$ (Fig.~\ref{fig:ResultsShear}(d); dashed lines). 
In the limit of $v_\text{max} \ll v_\textmd{s}$ we find the linear relation (Fig. \ref{fig:ResultsShear}(d); solid lines)
\begin{equation}
\label{eq:RetentionRatio}
R = \frac{2}{\pi} - \frac{2+9n +7n^2}{2+10n+12n^2} \frac{v_\text{max}}{v_\textmd{s}}.
\end{equation}
Hence, the difference in upstream retention ratio for shear-thinning and -thickening fluids grows with increasing flow speed. This determines the crossover between upstream or downstream motion of the majority of swimmers where $R=0$ (see inset of Fig.~\ref{fig:ResultsShear}(d)). 
The slopes change at larger flow speeds, $v_\text{max}>4 v_\textmd{s}$, when the tumbling trajectories start to outnumber the oscillating trajectories \cite{zottl2012nonlinear} and the full solution for $R$ must be applied (dashed lines).
In this $v_\textmd{max} \gg v_s$ regime, the difference in upstream retention ratio for shear-thinning and -thickening fluids can be large (Fig. \ref{fig:ResultsShear}(d)). 
For $v_\textmd{max} = 10 v_s$, the shear thickening ($n=2$) $R$-value differs by $33\%$ from the shear-thinning ($n=1/2$) value, which is substantial compared to the $5 - 10\%$ change in swimming speed observed in quiescent non-Newtonian fluids \cite{liu2011force, gagnon2014undulatory, gao2014altered, martinez2014flagellated, lauga2007propulsion, fu2010low, teran2010viscoelastic, spagnolie2013locomotion}.
\end{par}

\begin{par}
The significant modification of upstream retention ratios in non-Newtonian fluids can have important consequences in microbiological flows. For instance, our results suggest that a motile {\it H. pylori}, swimming with an average velocity of $27\mu m/s$ \cite{Celli2009} and subjected to gastric mucosal flow with a similar velocity and $n=0.5$, would have a $50\%$ reduction in upstream retention ratio than if it were swimming in a Newtonian fluid flow ($n=1$). 
Since the velocity of the mucosal flow can vary broadly~\cite{Sato1995} and $n$ can be as small as $\sim 0.15$~\cite{Sato1995,Rosler2012,Celli2009}, this serves as a conservative example.
\end{par}

\begin{par}
In addition to shear-dependent viscosities, many microbiological fluids are characterized by viscoelastic normal stress differences. 
To describe these, with a power-law viscosity, we employ the second-order fluid model \cite{Bird} with the stress tensor 
$ S_{ij} = -p \delta_{ij} + \eta(\dot{\gamma}) D^{(1)}_{ij} - \tfrac{1}{2} \psi_1 D^{(2)}_{ij}
 + (\psi_1 + \psi_2) D^{(1)}_{ik} D^{(1)}_{kj}$, 
where $\psi_1$ and $\psi_2$ are the first and second normal stress coefficients and
$D^{(1)}_{ij}$ and $D^{(2)}_{ij}$ are the Rivlin-Eriksen tensors.
The Deborah number is $\textmd{De} = \tfrac{\psi_1 - 2\psi_2}{\eta} \tfrac{v_\textmd{max}}{H} \ll 1$. 
The normal stress coefficients characterize the fluid elasticity. 
These terms do not alter the undisturbed flow profile of Eq. (3) in the absence of swimmers. 
However, the disturbance flow around a finite-sized swimmer in combination with non-uniform shear across the channel results in a normal stress imbalance that causes a lateral migration across streamlines. 
Normal stress-induced migration of passive, inertialess particles in pressure-driven flow is well documented \cite{ho1976migration, chan1977note, leal1980particle, LeshanskyPRL2007, d2010viscoelasticity, villone2011numerical, villone2011simulations, ardekani2012emergence, d2015particle}.
To determine the migration velocity we use Chan and Leal's solution for general quadratic flow \cite{chan1977note} by expanding the background flow profile (Eq.~\ref{eq:TextPoiseuilleFlow}) about the swimmer position, as reported previously~\cite{ardekani2012emergence}. In our system, the migration velocity is then 
\begin{eqnarray}
\label{eq:PoiseuilleMigrationVelocity}
\boldsymbol{v}_\textmd{M} = - \psi_n \left( \frac{|x|}{H} \right)^\frac{3-2n}{n} \boldsymbol{\hat{e}}_x, 
\end{eqnarray}
where $\psi_n = \psi_s a^2 v_\textmd{max}^{3-n} \gamma_0^{n-1} f(n) / \eta_0 H^{4-n}$, $f(n) = 5(1+n)^{3-n}/36 n^{4-n}$ and $\psi_s = \psi_1 - 2\psi_2$. 
The function $\psi_n$ encapsulates both the non-Newtonian effects of normal stress differences and shear-dependent viscosity. A viscoelastic torque $\boldsymbol{\Omega}_\textmd{M}$ is not included in Eq.~\eqref{eq:EquationsOfMotion2}~\cite{ardekani2012emergence} because this term is not significant compared to the vorticity when $\textmd{De} \ll 1$ and, by symmetry of the swimmer, does not lead to preferred orientations.
\end{par}

\begin{figure}[b]
	\includegraphics[width=\linewidth]{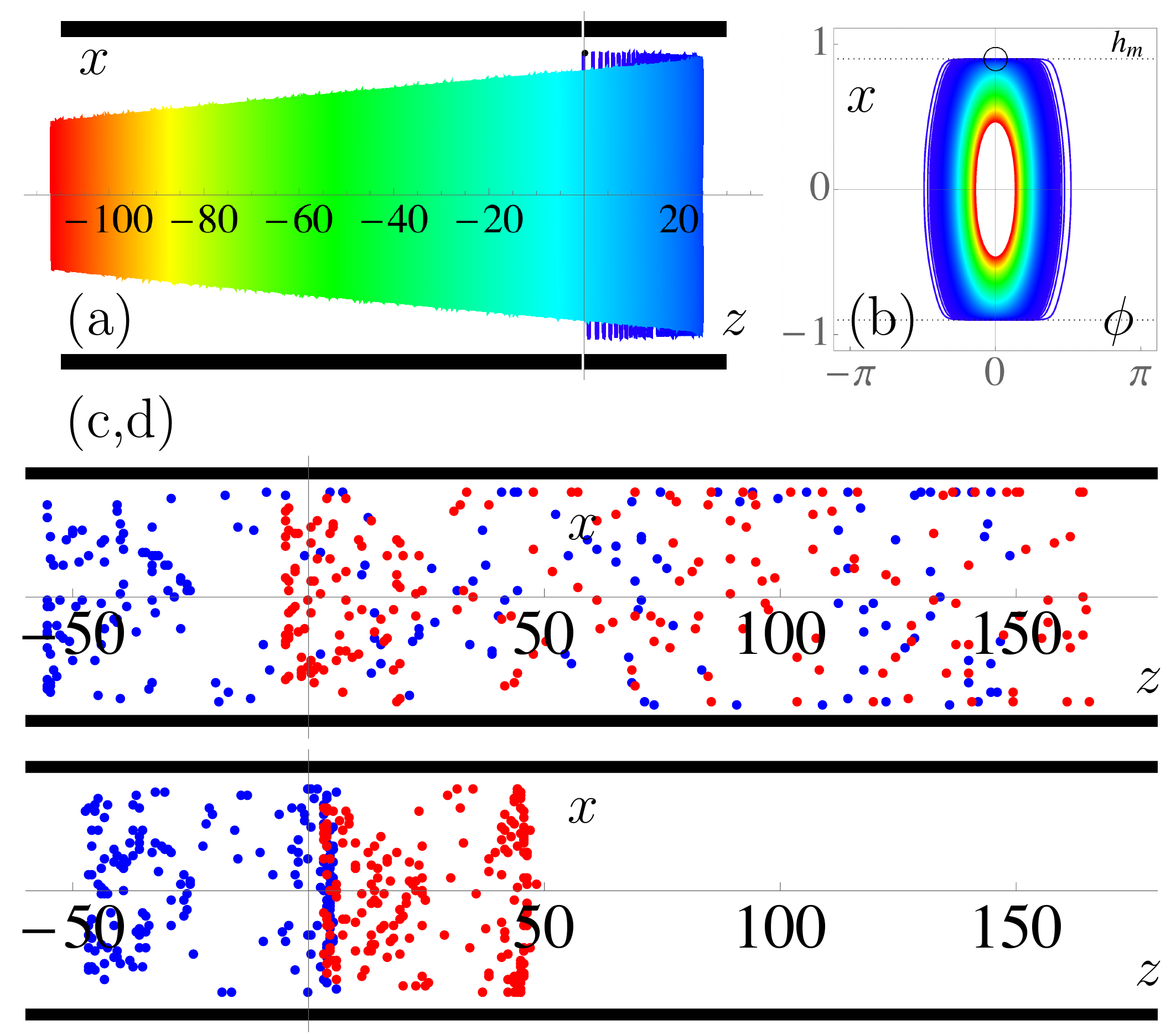}
        \caption{
Swimmer dynamics in Poiseuille flow of a shear-thinning viscoelastic fluid with $De = 0.1$, $n=0.8$, and $v_\textmd{max} = 1 v_\textmd{s}$. 
The swimmer parameters are $a=0.1$, $v_\textmd{s} = 1$, $\kappa = 0$ and $\sigma = 10^{-3}$.
a) Oscillating trajectory in the $x-z$ plane with initial position $\boldsymbol{r}_s=(0, 0.9)$ and orientations $\phi = 0$.
b) Corresponding trajectory in $x-\phi$ phase space. Colours indicate time progressing, from $t=0$ (blue) to $t=1000$ (red).
The swimmer is focussed towards the centerline and are reoriented to move upstream.
c-d) Two ensembles of swimmers, with $v_\textmd{s} = 1.1$ (blue) and $v_\textmd{s} = 0.9$ (red), are released from a random $x$-position and orientation in the channel at $z=0$. In a Newtonian fluid (c), the swimmers are dispersed but in a viscoelastic fluid (d) they remain clustered and are sorted according to swimming speed over time.
}
        \label{fig:ResultsVisco}
\end{figure}

\begin{par}
In both the shear-thinning and -thickening cases with normal stresses, the swimmer is driven to the centreline, and the coupling between motility and streamline migration rotates the swimmer to move upstream along the centreline (Fig.~\ref{fig:ResultsVisco}(a)). 
Unlike in a Newtonian fluid, the oscillations about the centreline are now damped in amplitude as the phase space origin ($x=\phi=0$) is a stable, attractive spiral (Fig.~\ref{fig:ResultsVisco}(b)).
The attraction is stronger for shear-thinning than shear-thickening fluids.
\end{par}

\begin{par}
We analyse this effect by linearising the equations of motion~(\ref{eq:EquationsOfMotion}-\ref{eq:EquationsOfMotion2}) about the origin so that $(\dot{\phi}, \dot{x})^T = \boldsymbol{M} (\phi, x)^T$ where
\begin{eqnarray}
\label{eq:EquationsOfMotionOrigin}
\boldsymbol{M} &=&
\left( \begin{array}{cc} -\frac{3 \kappa}{4} & v_\textmd{f} + \frac{3 \sigma }{2}
\\ -v_\textmd{s} + \frac{\sigma}{4} + \frac{3\nu}{2} & \frac{3 \kappa }{2} - \psi_n\end{array} \right). 
\end{eqnarray}
In $\boldsymbol{M}$, $\psi_n$ and dipolar HI terms are responsible for the spiral. 
Away from the walls, viscoelasticity dominates over HI effects and the eigenvalues of $\boldsymbol{M}$ without HI are found to be $\lambda_\pm = \tfrac{1}{2}(-\psi_n \pm \sqrt{\psi_n^2 - 4 v_\textmd{max} v_\textmd{s}})$.
Hence, the origin is a stable fixed point if $\psi_n^2 > 4 v_\textmd{max} v_\textmd{s}$ with two real negative (attractive) eigenvalues. 
Otherwise, the origin is a stable spiral with complex eigenvalues and negative real parts, meaning that swimmers perform damped oscillations about the centreline as verified in Fig.~\ref{fig:ResultsVisco}(a,b). 
Because the function $f(n)$ decreases monotonically with $n$, $\psi_n$ is larger for shear-thinning fluids and therefore the attraction towards the centreline is greater. 
\end{par}

\begin{par}
Though more pronounced in shear-thinning than shear-thickening flows, swimmers in flowing viscoelastic fluids tend to move upstream along the centreline after some time, regardless of initial position or upstream orientation. 
This allows for a sorting mechanism to select swimmers with a given swimming speed larger than the tunable Poiseuille flow, as demonstrated in  Fig.~\ref{fig:ResultsVisco}(c,d), where distributions of swimmers with different self-propulsion velocities are initially introduced at random positions and orientations in the channel in Newtonian and shear-thinning viscoelastic fluids.  
Unlike the Newtonian fluid, swimmers with larger motility are separated by moving upstream in the viscoelastic fluid (see Supplemental Material movies 2,3 in Ref. \cite{Movies}). 
It is worth noting that we expect this sorting mechanism to be robust against translational and orientational noise since small amounts of noise will keep the oscillation size nonzero, enhancing the upstream retention ratio and hence the sorting. 
\end{par}

To summarize, unlike the prevalent boundary accumulation in quiescent Newtonian fluids, swimmers' trajectories show oscillatory motion about the centreline. 
Average migration against Poiseuille flows is enhanced (reduced) in shear-thickening (-thinning) fluids compared to simple Newtonian fluids. 
It is not necessary that the non-Newtonian nature of these fluids be appreciable on the microscale since altered trajectories arise from differences in vorticity at macroscopic scales. 
This constitutes a substantial change to the effective upstream motility, that is comparable to or greater than observed changes in motility due to microscopic effects on swimming in quiescent non-Newtonian fluids \cite{liu2011force, gagnon2014undulatory, gao2014altered, martinez2014flagellated, lauga2007propulsion, fu2010low, teran2010viscoelastic, spagnolie2013locomotion}.
The oscillations are damped towards the centreline in the presence of viscoelastic normal stress differences resulting in direct upstream migration. 
This offers a sorting mechanism to differentiate motile microorganisms according to speed.

\begin{acknowledgments}
We gratefully thank Andreas Z\"{o}ttl for helpful comments on the manuscript.
This work was funded through ERC Advanced Grant (291234 MiCE), and we acknowledge EMBO for support to T.N.S (ALTF181-2013).
\end{acknowledgments} 

\bibliography{lit.bib}

\end{document}